\documentstyle[prl,aps,twocolumn,epsf]{revtex}

\begin{document}

\draft

\title{Shape deformations and angular momentum transfer \\
       in trapped Bose-Einstein condensates}

\author{F. Dalfovo$^1$ and S. Stringari$^2$}

\address{$^1$ Dipartimento di Matematica e Fisica, Universit\`{a} 
Cattolica, Via Musei 41, Brescia, Italy \\
and Istituto Nazionale per la Fisica della Materia,
Unit\`{a} di Brescia}

\address{$^2$ Dipartimento di Fisica, Universit\`{a} di Trento, 
I-38050 Povo, Italy \\
and Istituto Nazionale per la Fisica della Materia, Unit\`{a} 
di Trento}

\date{September 13, 2000}

\maketitle

\begin{abstract}
Angular momentum can be transferred to a trapped Bose-Einstein condensate by 
distorting its shape with an external rotating field, provided the rotational 
frequency is larger than a critical frequency fixed by the energy and angular 
momentum of the excited states of the system. By using the Gross-Pitaevskii 
equation and sum rules, we explore the dependence of such a critical frequency 
on the multipolarity of the excitations and the asymmetry of the confining 
potential. We also discuss its possible relevance for vortex nucleation in 
rotating traps.  
\end{abstract}

\pacs{PACS numbers: 03.65.-w, 05.30.jp, 32.80.-t, 67.40.Db}


It is well known that in order to create excitations in a superfluid by 
moving an impurity one has to overcome some critical velocity, which is 
fixed by the spectrum of the excited states of the system. In the case of a
rectilinear motion of a heavy object, conservation of energy and 
momentum yields the critical velocity $v_c = {\rm min}[ \epsilon(p)/p ]$, 
where $\epsilon(p)$ is the energy of an excitations carrying momentum $p~$;  
this is known as Landau criterion \cite{Landau}. The external perturbation
can also be a moving boundary, like the wall of a container, which produces 
excitations through its roughness. In liquid helium such a critical velocity 
has been the object of a longstanding investigation, involving the
structure of the phonon-roton branch and the nucleation and dynamics of 
quantized vortex lines and rings. The 
existence of similar critical velocities in trapped Bose-Einstein 
condensates of alkali atoms is currently under study. In 
Ref.~\cite{MIT-Landau} an external potential (a laser beam) is used to 
produce a hole in the condensate and this hole is moved back and forth 
at variable velocity, playing the role of a massive impurity. Theoretical
interpretations in terms of the Landau criterion have been already 
presented in Refs.~\cite{adams,shlya}. Here we consider the analog process 
in which the perturbation acts at the surface of the condensate, producing 
a rotating 
shape deformation. This kind of perturbation has been recently 
implemented in \cite{MIT-shape} to excite quadrupole and octopole 
modes, but the analysis is also relevant for rotating traps, like
the one used in Ref.~\cite{ENS} to produce quantized vortices. 
The similarity with liquid helium is again worth stressing; in that 
case, when the liquid fills a rotating bucket, the roughness of the 
wall can transfer angular momentum to the superfluid, favoring the 
occurrence of quantized vorticity above some critical angular
velocity \cite{helium}. However, since the crucial length scales 
(healing length, surface thickness, interatomic distance, etc.) 
in liquid helium and trapped condensates are different, the role of 
surface excitations on the Landau criterion is also very different. 
In particular, surface modes can be ignored in the analysis of the 
experiments with helium in rotating vessels, while they might be 
crucial for gases of alkali atoms in rotating traps, as discussed below. 

Let us consider a condensate which is initially in its ground state in
an axially symmetric trap, with trapping frequencies $\omega_\perp$ and 
$\omega_z=\lambda \omega_\perp$, where $\lambda$ is the anisotropy 
parameter of the harmonic confining potential. 
Then, let us imagine a small distortion of the condensate caused by an 
external perturbation which rotates around 
the $z$-axis at angular velocity $\Omega$.  This perturbation may be a 
modulation of the confining potential, a rotating laser beam, or 
something else; we only assume here that the consequent distortion of the 
condensate is indeed a shape deformation, coupled to the surface modes of 
the system (ripplon-like waves). Let us consider the process in which 
the external probe produces a single excited state, corresponding to a 
surface wave with energy $\hbar \omega(\ell)$ and angular momentum $\ell$ 
along $z$ (azimuthal quantum number $m=\ell$). One can assign a
moment of inertia $I$ to the rotating perturbation and write the 
conservation laws for energy and angular momentum, letting $I \to \infty$ 
at the end. The simple result is that the conservation laws are both 
satisfied only if $\Omega$ is larger than the critical angular velocity 
$\Omega_c= {\rm min}[\omega(\ell)/\ell]$. This is the analog of the 
Landau criterion.  If the rotation is slower than 
$\Omega_c$, the condensate simply adjusts its shape to the rotating 
perturbation by moving as an irrotational and nondissipative fluid. 
Conversely, if the rotation is faster than $\Omega_c$, the external 
perturbation can excite surface modes in a continuous way, transferring 
energy and angular momentum to the condensate.  

The excitation spectrum of a condensate can be accurately evaluated by 
solving the linearized Gross-Pitaevskii (GP) equation, also known as 
Bogoliubov's equations \cite{RMP}. The theory provides the energy of 
the excited states at zero temperature in terms of the relevant quantum 
numbers (number of radial nodes and angular momentum). For our 
purposes what matters is the energy $\hbar \omega (\ell)$ of the surface 
modes, i.e., the ones with no radial nodes. In a previous work, we already 
presented some results for the critical angular velocity $\Omega_c$ in a 
spherical condensate \cite{Muntsa}.  Here we discuss more systematically
the consequences of this critical behavior  and provide results for  
nonspherical traps.  

We avoid the complication of solving the Bogoliubov's equations for 
axially symmetric condensates by using sum rules. As already shown in 
\cite{Stringari96}, the sum rule approach provides rigorous upper bounds 
to the dispersion $\omega(\ell)$ using the ground state density, solution 
of the stationary GP equation, as main ingredient. The key quantities are 
the $k$-energy weighted moments, $m_k= \int_0^{\infty} S_F(E) E^k dE$ of 
the dynamic structure factor, 
$S_F(E) = \sum_j | \langle j | F |0 \rangle |^2 \delta ( E - E_{j0}) $, 
associated with a given excitation operator $F$, where $E_{j0}= (E_j-E_0)$ 
is the excitation energy of the eigenstate $|j\rangle$ of the 
Hamiltonian. The $m_1$ and $m_3$ moments can be written as expectation
values of commutators,  $m_1= (1/2) \langle 0 | [ F^\dagger , 
[H, F]] |0 \rangle $ and $m_3= (1/2) \langle 0 | [[F^\dagger ,H], [H,  
[H, F]]] |0 \rangle $, and the ratio $(m_3/m_1)^{1/2}$ gives the desired
upper bound to the energy of the states excited by the operator $F$. In 
order to select the surface modes, one  chooses the excitation operator 
in the form $F=\sum_{i=1}^N r_{i}^\ell Y_{\ell \ell} (\theta_i,\phi_i)$,
with $\ell \ne 0$, where $N$ is the number of atoms in the condensate. 
The explicit calculation of the ratio $(m_3/m_1)$ with this 
operator gives the dispersion law
\begin{equation}
\omega^2 (\ell) = \omega_\perp^2 \ell [ 1 + (\ell -1) \beta_\ell ] \; , 
\label{eq:m3overm1}
\end{equation}
where 
\begin{equation}
\beta_\ell = \frac{ \int d{\bf r} \ ( \nabla_\perp \sqrt{n} )^2 
r_\perp^{2\ell -4} }{ \int d{\bf r} \ n \ r_\perp^{2\ell-2} } 
\label{eq:beta}
\end{equation}
and $r_\perp = (x^2+y^2)^{1/2}$. Here $n(r_\perp,z)$ is the ground state 
density of the condensate which
 can be calculated by solving the stationary 
GP equation.  The sum rule estimate is close to the 
exact Bogoliubov's spectrum whenever the state excited by $F$
is highly collective, with a strength which almost exhausts the dynamic
structure factor. This is true for shape deformations of low multipolarity,
which are the important ones for the present analysis. The dynamic structure
factor associated with large $\ell$ excitations is expected to be more and 
more fragmented and, consequently, the sum rule upper bound becomes less  
accurate. 

The surface state with lowest energy is the quadrupole mode. The sum rule 
estimate is easily obtained by inserting $\ell=2$ into 
Eqs.~(\ref{eq:m3overm1})-(\ref{eq:beta}). One finds
\begin{equation}
\omega^2 (\ell=2) = 2 \omega_\perp^2 \left( 1 +
{ (E_{\rm kin})_\perp \over  (E_{\rm ho})_\perp } \right)
\label{eq:quadrupole}
\end{equation}
where $(E_{\rm kin})_\perp$ and $(E_{\rm ho})_\perp$ are the radial 
contributions to the kinetic and external potential energies, 
respectively. For noninteracting particles these two energies are equal
and one gets $\omega = 2 \omega_\perp$. The other interesting limit 
is $N \to \infty$ (Thomas-Fermi limit), when the kinetic energy in the
ground state is much smaller than both the atom-atom interaction energy and
the trapping potential. In this limit Eq.~(\ref{eq:quadrupole}) gives
$\omega = \sqrt{2} \ \omega_\perp$. The latter result can also be obtained 
by solving the hydrodynamic equations of a superfluid \cite{Stringari96}, 
which provide the general dispersion law $\omega(\ell)=\sqrt{\ell} 
\ \omega_\perp$ for the surface modes with $m=\pm \ell$. The hydrodynamic
approach, however, is accurate only for the lowest multipoles when one
can completely neglect the contribution of the quantum pressure term 
in the GP equation; for the values of $N$ which are relevant in the current 
experiments, the hydrodynamic dispersion becomes soon inaccurate as $\ell$ 
increases. This has an important consequence for the evaluation of the
critical frequency $\Omega_c= {\rm min}[\omega(\ell)/\ell]$, since the 
hydrodynamic dispersion gives $\omega(\ell)/\ell=\omega_\perp/ \sqrt{\ell}$ 
and hence $\Omega_c=0$; conversely, $\Omega_c$ always exhibits a minimum, 
even for very large $N$, when calculated with Bogoliubov's equations or 
using the sum rule results (\ref{eq:m3overm1})-(\ref{eq:beta}).

\begin{figure}[t]
\epsfysize=8cm
\hspace{3cm}
\epsffile{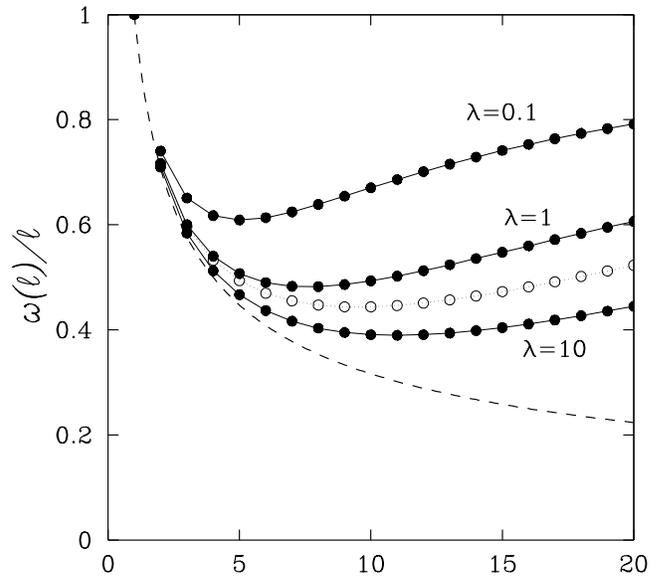}
\caption{ The quantity $\omega(\ell)/\ell$, in units of $\omega_\perp $,
obtained with Eqs.~(\protect\ref{eq:m3overm1})-(\protect\ref{eq:beta})
as a function of $\ell $ for three condensates of $^{87}$Rb atoms
with different values of $\lambda=\omega_z/\omega_\perp$: prolate 
($\lambda=0.1$), spherical ($\lambda=1$), oblate ($\lambda=10$).  The 
three condensates have the same value of $N=10^5$ and 
$\omega_{\rm ho}=\lambda^{1/3} \omega_\perp = 2 \pi (20 {\rm Hz} )$. 
Open circles correspond  to the exact solution of the 
Bogoliubov's equation in the spherical case ($\lambda=1$). 
The dashed line is the 
hydrodynamic prediction $\omega(\ell)/\ell=1/\protect\sqrt{\ell}$. }
\label{fig1}
\end{figure}

\begin{figure}[t]
\epsfysize=8cm
\hspace{3cm}
\epsffile{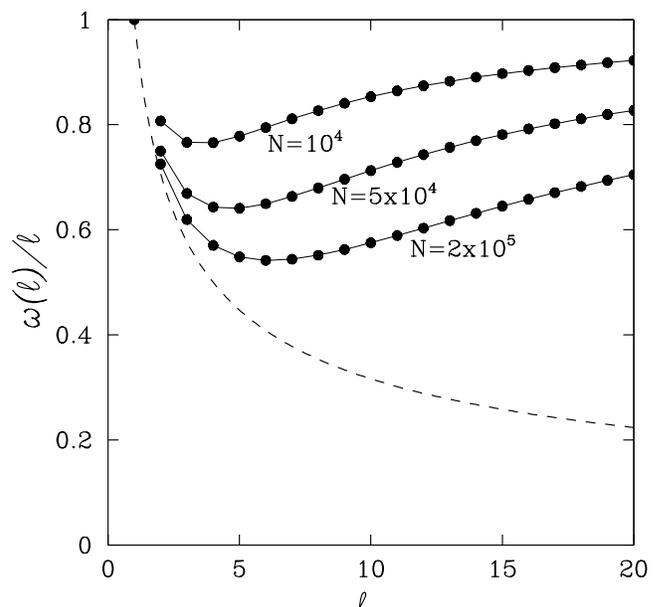}
\caption{ Same as in Fig.~1 but for condensates with different values of
$N$. All condensates have the same $\omega_{\rm ho}= 2 \pi (82.5 {\rm Hz})$ 
and $\lambda=0.06$. The parameters used for the curve $N=2\times 10^5$ are 
close to the actual situation in the experiments of Ref.~\protect\cite{ENS}.
}
\label{fig2}
\end{figure}

In Fig.~\ref{fig1} we show typical results for $\omega(\ell)/\ell$, 
in units of $\omega_\perp$, obtained by inserting in 
Eqs.~(\ref{eq:m3overm1})-(\ref{eq:beta}) the ground state density 
of three different condensates of $^{87}$Rb atoms (we use $a=100 a_0$ 
for the scattering length, where $a_0$ is the Bohr radius). The stationary 
GP equation has been solved numerically as in Ref.~\cite{Dalfovo}. 
The condensates have the same number  
of atoms, $N=10^5$, and the same average trapping frequency, 
$\omega_{\rm ho} = (\omega_x \omega_y \omega_z)^{1/3}= \lambda^{1/3} 
\omega_\perp = 2\pi (20 {\rm Hz})$, but different asymmetry 
parameter: $\lambda=0.1$ (prolate), $\lambda=1$ (spherical) and 
$\lambda=10$ (oblate). The $N$-dependence of $\omega(\ell)/ \ell$ is 
instead given in Fig.~\ref{fig2}, where we plot the curves 
for condensates with different $N$ but the same prolate geometry: 
$\lambda=0.06$ and $\omega_{\rm ho} = 2\pi (82.5 {\rm Hz})$, close to 
the values of Ref.~\cite{ENS}.  In order to test the accuracy 
of the sum rule approach, we have also solved numerically the 
Bogoliubov's equations for a spherical condensate. The results are 
shown as open circles in Fig.~\ref{fig1} and can be compared with the 
sum rule results for $\lambda=1$ (solid circles). As one can see, the 
accuracy of the sum rule approach is very good for the lowest values 
of $\ell$, remaining of the order of $10$\% \ up to $\ell=10$. This 
is enough for our purposes.  Finally, the hydrodynamic prediction is 
also shown as a dashed line in both figures.

In Table I we summarize the situation for three different condensates. 
The parameters of the first (prolate) and the second (spherical) are close 
to the ones of current experiments at ENS-Paris \cite{ENS} and JILA-Boulder 
\cite{JILA}, respectively. The third one (oblate) is presented in order to 
stress the dependence of the critical frequencies on the geometry of
the trap. Actually, the critical frequencies $\omega(\ell)/\ell$ for the
lowest multipoles ($\ell=2,3,4$) turn out to be rather independent of
the shape of the condensate and are well described by the hydrodynamic
dispersion, which is independent of $\lambda$. For the quadrupole  mode,
for example, it is very well reproduced by the hydrodynamic value
$\omega_\perp/\sqrt{2}=0.707 \omega_\perp$. The minimum of $\omega(\ell)
/\ell$ instead depends on $\lambda$ in a more significant way, since it 
involves higher $\ell$'s, especially for the oblate geometry. 

The frequencies $\omega(\ell)/\ell$ in Table I are thresholds for the 
transfer of angular momentum to a condensate through shape deformations.  
The values of $\ell$ actually involved in a given experiment depend on the 
type of process used.  For instance, a laser beam which makes a rotating 
hollow on the surface of the condensate might excite a bunch of states with 
several $\ell$'s, just as happens for a hole moving through
the fluid and exciting phonon-like excitations with several momentum
components.  
Conversely, the stirred confining potential of Ref.\cite{ENS}, with a 
rotating quadrupole deformation, is expected to excite mostly quadrupole 
excitations. What really happens just above, or close to the threshold can 
not be described only in terms of elementary excitations. 
In fact, once a threshold is reached, the rotating perturbation can pump  
many excitations into the condensate, bringing it far from the linear regime. 
A wider configuration space is made available and the system can find 
new rotating equilibrium states, like the stationary deformed configurations
discussed in \cite{recati}, or jump into states with quantized
vorticity. 

Concerning vortices, an important frequency scale is fixed by the 
difference between the energy per particle of a condensate with and 
without vortex calculated in the rotating frame,  $\Omega_v = 
(1/\hbar) [ (E/N)_v - (E/N)_g]$. This corresponds to the ratio 
between the `excitation' energy of the vortex in the rest frame, 
$(E_v-E_g)$, and its angular momentum $N\hbar$.  When the rotation 
frequency $\Omega$ is equal to $\Omega_v$ the energies of the 
configurations with and without vortex are equal in the rotating
frame and, above $\Omega_v$, the vortex state becomes the global energy 
minimum.  In this sense, $\Omega_v$ is analog to $\omega(\ell)/\ell$, 
since the latter represents the frequency above which a surface state 
of angular momentum $\ell$ has negative energy in the rotating 
frame \cite{machida}. 
A crucial difference, however, is that in order to transfer $N\hbar$
angular momentum and nucleate a vortex above $\Omega_v$ one has to 
overcome an energy barrier, while no barrier is found above $\omega(\ell)
/\ell$ for the creation of quanta of angular momentum through 
surface excitations.  

The quantity $\Omega_v$ can be obtained from 
the solution of the stationary Gross-Pitaevskii equation for the ground 
state and the state with a vortex line along $z$. We have already presented 
some numerical results for this quantity in \cite{Dalfovo}. By repeating 
the same kind of calculation for the three condensates in Table I we find
the values given in the last column \cite{Lundh}. It has been noted in 
Ref.~\cite{ENS} that the experimental result for the frequency at which 
vortices occur is significantly larger than $\Omega_v$. The critical 
frequency in the experiment turns out to be about $0.67 \omega_\perp$, to 
be compared to the value $0.35 \omega_\perp$ given in the first row of 
Table I. This is likely to be the consequence of the above mentioned
energy barrier, which prevents the nucleation of vortices even above
$\Omega_v$.  If one tries to impart a rotation to the condensate by 
rotating a slightly asymmetric external field, the crucial question is 
how angular momentum is actually transferred to the system. Our 
analysis suggests that, if the transfer mechanism is the creation of 
surface excitations of multipolarity $\ell$, the relevant threshold is 
$\omega(\ell)/\ell$, which can be significantly larger than 
$\Omega_v$, as shown in Table I. This argument might be 
used to interpret the fact that 
the experimental critical frequency of Ref.~\cite{ENS} turns out to be
closer to the quadrupole value $\omega(2)/2 \sim \omega_\perp/\sqrt{2}$ 
than to $\Omega_v$. Were this true, one should find almost the 
same critical frequency by repeating the experiment in spherical or 
oblate condensates. This clearly suggest that the $\lambda$-dependence 
could discriminate between alternative descriptions for 
vortex nucleation and stability (see, for instance, 
Refs.~\cite{fetter,clark,garcia} and references therein). 
Another important
signature of the suggested threshold mechanism would be the observation 
of strongly deformed configurations in the initial stage of the vortex 
nucleation. Preliminary evidence for such deformations has already
emerged \cite{dalibard}. Finally, the threshold 
behavior here discussed would imply a significant hysteresis when the 
rotation frequency is first speeded up to produce vortices
and then slowed down again, since one expects different mechanisms
for the two processes: the nucleation of the vortex state at relatively
high rotation frequency and the destabilization of the vortex at
lower frequency (for an extensive discussion about the destabilization 
of vortices see \cite{fetter}). 
The hysteresis mechanism of the vortical configurations has some 
similarity with the behavior of superfluid helium.  In that case, 
however, vortex nucleation is associated with the pinning of vortex lines 
at the walls of the container \cite{helium}. Trapped condensates 
have no rough surfaces or, in other words, they have an adjustable 
``roughness"; in this sense, they offer a novel and
complementary tool for understanding nucleation processes
of quantized vorticity in superfluids.

This work has been supported by the Ministero dell'Universit\`{a} e 
della Ricerca Scientifica e Tecnologica. F.D. would like to thank the 
Dipartimento di Fisica di Trento for hospitality.



\begin{center}
\begin{tabular} {||c||c|c|c|c||c||}
\hline
Condensate & 
$\frac{\omega(2)}{2}$ & 
$\frac{\omega(3)}{3}$ & 
$\frac{\omega(4)}{4}$ & 
${\rm min}[\frac{\omega(\ell)}{\ell}]$  & 
$\Omega_v$  \\
\hline 
\hline
prolate              & $0.72$ & $0.61$ & $0.56$ & $0.53$ & $0.35$    \\
$\lambda=0.0058$                       & & & & &        \\
$\nu_\perp = 175 {\rm Hz}$  & & & & &        \\
$N=2.5 \times 10^5$                    & & & & &        \\ 
\hline
spherical            & $0.71$ & $0.59$ & $0.53$ & $0.44$ & $0.24$    \\
$\lambda=1$          & $(0.71)$ & $(0.59)$ & $(0.52)$ & $(0.41)$ &    \\
$\nu_\perp = 7.8 {\rm Hz}$  & & & & &        \\
$N=3 \times 10^5$                      & & & & &        \\
\hline
oblate               & $0.71$ & $0.58$ & $0.50$ & $0.33$ & $0.12$    \\
$\lambda=10$                           & & & & &        \\
$\nu_\perp = 10 {\rm Hz}$   & & & & &        \\
$N=3 \times 10^5$                      & & & & &        \\
\hline
\hline
\end{tabular}
\end{center}
{\bf Table I}. Critical frequencies for shape deformations in different
condensates. First column: type of condensate. Next four columns: 
critical frequency $\omega(\ell)/\ell$, in units of the radial
trapping frequency $\omega_\perp = 2\pi \nu_\perp$, obtained with the sum 
rule expressions (\protect\ref{eq:m3overm1})-(\protect\ref{eq:beta}) for 
$\ell=2,3,4$ and for its minimum value. The numbers in brackets are the 
exact results of Bogoliubov's equations in the spherical case. The last 
column is the quantity  
$\Omega_v = (1/\hbar)[ (E/N)_v - (E/N)_g]$ (see text). 

\begin{references}


\bibitem{Landau} L.D. Landau, J. Phys. (USSR) {\bf 5}, 71 (1941).

\bibitem{MIT-Landau} C. Raman {\it et al.}, Phys. Rev. Lett. {\bf 83},
2502 (1999).

\bibitem{adams} B. Jackson, J.F. McCann, and C.S. Adams, Phys. Rev.
A {\bf 61}, 051603(R) (2000); T. Winiecki {\it et al.}, 
e-print cond-mat/0004430.

\bibitem{shlya} P.O. Fedichev and G.V. Shlyapnikov, e-print
 cond-mat/0004039.

\bibitem{MIT-shape} R. Onofrio {\it et al.}, Phys. Rev. Lett. {\bf 84},
810 (2000)

\bibitem{ENS} K.W. Madison, F. Chevy, W. Wohlleben, and J. 
Dalibard, Phys. Rev. Lett. {\bf 84}, 806 (2000); F. Chevy, K.W. 
Madison and J. Dalibard, e-print cond-mat/0005221. 

\bibitem{helium} R.J. Donnelly, {\it Quantized Vortices in Helium II} 
(Cambridge University Press, Cambridge 1991). The pioneering 
papers by G.B. Hess and W.M. Fairbank, Phys. Rev. Lett. {\bf 19}, 
216 (1967) and by R.E. Packard and T.M. Sanders, Phys. Rev. A {\bf 6}, 
799 (1972) are also worth reading in view of the most recent analysis
done in Ref.~\protect\cite{ENS} on the angular momentum in trapped
condensates.  

\bibitem{RMP}  F. Dalfovo, S. Giorgini, L. Pitaevskii and S. Stringari, 
Rev. Mod. Phys. {\bf 71},
463 (1999).

\bibitem{Muntsa} F. Dalfovo {\it et al.},
Phys. Rev. A {\bf 56}, 3840 (1997).

\bibitem{Stringari96}  S. Stringari, Phys. Rev. Lett. {\bf 77}, 2360 (1996).

\bibitem{Dalfovo} F. Dalfovo and S. Stringari, Phys. Rev. A {\bf 53},
2477 (1996).

\bibitem{JILA} M.R. Matthews {\it et al.}, Phys. Rev. Lett. {\bf 83},
2498 (1999)

\bibitem{recati} A. Recati, F. Zambelli and S. Stringari, e-print
cond-mat/0007152.

\bibitem{machida} See also the discussion in 
T. Isoshima and K. Machida, Phys. Rev.
A {\bf 60}, 3313 (1999).

\bibitem{Lundh}  A simple analytic expression, valid in the Thomas-Fermi
limit, has been also derived by E. Lundh, C.J. Pethick and H. Smith,
Phys. Rev. A {\bf 55}, 2126 (1997). For the three values of $\Omega_v$ 
in Table I we checked that their approximate formula differs
from the exact GP result by less than $2$\%. 

\bibitem{fetter} A. Svidzinsky and A. Fetter, e-print 
cond-mat/0007139 and references therein.  

\bibitem{clark} D.L. Feder, C.W. Clark, and 
B.I. Schneider, Phys. Rev. Lett. {\bf 82}, 4956 (1999) and Phys. Rev. 
A {\bf 61}, 011601 (2000); D.L. Feder {\it et al.}, e-print
cond-mat/0009086.  

\bibitem{garcia} J.J. Garc\'{\i}a-Ripoll and V.M. P\'erez-Garc\'{\i}a,
e-print cond-mat/0006368. 

\bibitem{dalibard} J. Dalibard, private communication. 

\end{references}
\end{document}